\documentclass[twocolumn,prb,showpacs]{revtex4}%
\usepackage{graphicx}%
\usepackage{amsmath}%
\usepackage{amsfonts}%
\usepackage{amssymb}
\usepackage{bm}

\def\s{{\sigma}}
\def\e{{\epsilon}}
\def\k{{ {\bm k} }}

\def\q{{ {\bm q} }}
\def\Q{{ {\bm Q} }}

\def\w{{\omega}}
\def\a{{\alpha}}

\begin{document}
\title{Neutron-Inelastic-Scattering Peak 
by Dissipationless Mechanism \\
in the $s_{++}$-wave State in Iron-based Superconductors
}
\author{Seiichiro \textsc{Onari}$^{1}$
 and Hiroshi \textsc{Kontani}$^{2}$}
\date{\today }

\begin{abstract}
We investigate the neutron scattering spectrum in iron pnictides 
based on the random-phase approximation in the five-orbital model
with a realistic superconducting (SC) gap, $\Delta=5$meV.
In the normal state, the neutron spectrum is suppressed by 
large inelastic quasi-particle (QP) scattering rate $\gamma^*\sim\Delta$.
In the fully-gapped $s$-wave state
without sign reversal ($s_{++}$), a hump-shaped enhancement
appears in the neutron spectrum just above $2\Delta$,
since the inelastic QP scattering is prohibited by the SC gap.
That is, the hump structure is produced by
the dissipationless QPs for QP energy $E_\k<3\Delta$.
The obtained result is more consistent with experimental spectra,
compared to the results of our previous paper with $\Delta=50$meV.
On the other hand, both height and weight of the resonance peak
in the fully-gapped $s$-wave states with sign reversal ($s_{\pm}$) 
are much larger than those observed in experiments.
We conclude that experimentally observed 
broad spectral peak in iron pnictides is created by the present
``dissipationless mechanism'' in the $s_{++}$-wave state. 
\end{abstract}

\address{
$^1$ Department of Applied Physics, Nagoya University and JST, TRIP, 
Furo-cho, Nagoya 464-8602, Japan. 
\\
$^2$ Department of Physics, Nagoya University and JST, TRIP, 
Furo-cho, Nagoya 464-8602, Japan. 
}
 

\sloppy

\maketitle

\section{Introduction}
Since the discovery of superconductivity in iron pnictides with high
transition temperature ($T_c$)\cite{Hosono}, substantial experimental and
theoretical works have been performed to clarify the mechanism of
superconductivity.
The superconducting (SC) gap in many iron pnictides is fully-gapped and
band-dependent, as shown by the penetration depth measurement \cite{Matsuda}
and the angle-resolved photoemission spectroscopy (ARPES)
 \cite{ARPES1,ARPES4}.
The fully-gapped state is also supported by the rapid suppression 
in $1/T_1$ ($\propto T^{n}$; $n\sim4-6$) below $T_{\rm c}$ 
\cite{Sato-T1,Mukuda,Grafe}.
On the other hand, P-doped Ba122 \cite{AsP} and
LaFePO\cite{LaFePO,LaFePO-penetration} show the nodal line behavior
($T$-linear dependence) in penetration depth measurements. In these
compounds, $A_{1g}$ symmetry pairing states with accidental nodes are expected
theoretically.\cite{graser,onari-flex} 

In iron pnictides, the intra-orbital nesting of the Fermi surface (FS)
between the hole- and electron-pockets is expected to induce
the antiferromagnetic (AF) fluctuations.
Taking this fact into account, the fully-gapped sign-reversing
$s$-wave state ($s_\pm$-wave state) mediated by the AF fluctuation
had been predicted\cite{Kuroki,Mazin}.
On the other hand, we have demonstrated that the orbital fluctuation
mediated fully-gapped $s$-wave state 
without sign-reversal ($s_{++}$-wave state) is realized by
the inter-orbital nesting, by taking the electron-phonon interaction into
account.\cite{Kontani-Onari,Saito}
In the latter scenario, 
the close relation between $T_{\rm c}$ and the 
crystal structure revealed by Lee \cite{Lee-plot}, 
{\it e.g.}, $T_{\rm c}$ becomes the highest
when the As$_4$ cluster is regular tetrahedron,
is automatically explained \cite{Saito}.
Moreover, the latter scenario is 
consistent with the large SC gap on the $z^2$-orbital band 
in Ba122 systems \cite{Saito}, 
observed by bulk-sensitive laser ARPES measurement \cite{Shimo-Science}.
In addition, the orthorhombic structure transition and the corresponding
shear modulus softening is well explained theoretically \cite{Softening}.
In newly discovered K$_x$Fe$_2$Se$_2$ with $T_c\sim 30$K,
in which only electron-pockets exist,
orbital-fluctuation-mediated fully-gapped $s_{++}$-wave state\cite{saito-Se} or
spin-fluctuation-mediated nodal $d_{x^2-y^2}$-wave state\cite{saito-Se,hirschfeld-imp} had been predicted theoretically.

Thus, it is important to clarify the sign of the SC gap
via phase-sensitive experiments.
Nonmagnetic impurity effect offers us useful phase-sensitive information.
In iron pnictides, the SC state survives against high substitution of
Fe sites by other element (more than 10\%).
 \cite{Kawabata,Co,Ru,122-phase}
These results support the $s_{++}$-wave state
since the $s_{\pm}$-wave state is very fragile against impurities,
similar to other unconventional superconductors.
\cite{onari-impurity,kontani-sato}
Moreover, impurity driven crossover from $s_{\pm}$-wave state to
$s_{++}$-wave state had been
discussed in Refs.\cite{onari-flex,hirschfeld-imp}.

Another promising method is the neutron scattering measurement:
As discussed by Monthoux and Scalapino in Ref. \cite{Monthoux-Scalapino}, existence of the resonance peak at a nesting wavevector $\bm{Q}$
is a strong evidence for AF fluctuation mediated 
superconductors with sign reversal
\cite{Monthoux-Scalapino,pines,chubukov-resonance,takimoto-moriya}.
The resonance occurs under the condition $\w_{\rm res}<2\Delta$,
where $\w_{\rm res}$ is the resonance energy and 
$\Delta$ is magnitude of the SC gap.
The sharp and large resonance peak has been observed in many
AF fluctuation mediated unconventional superconductors, like
high-$T_c$ cuprates \cite{iikubo-sato,ito-sato,keimer-highTc},
CeCoIn$_5$ \cite{stock-CeCoIn5}, and UPd$_2$Al$_3$ \cite{sato-UPd2Al3}. 
The measurements of phonon spectral function for $|\w|\lesssim 2\Delta$ 
would also be useful.\cite{scalapino}

Neutron scattering measurements for iron pnictides had been performed 
\cite{christianson,keimer,zhao,qiu,tate,Sato-neutron}
after the theoretical predictions \cite{maier-scalapino,eremin}.
Although clear peak structures were observed in
FeSe$_{0.4}$Te$_{0.6}$ \cite{qiu}, BaFe$_{2-x}$Co$_{x}$As$_2$ \cite{keimer,tate}
and Ca-Fe-Pt-As \cite{Sato-neutron},
these weights are much smaller than those in high-$T_c$ cuprates and
CeCoIn$_5$. 
Moreover, the resonance condition $\w_{\rm res}<2\Delta$
is not surely confirmed since it is difficult to determine the value of
$\Delta$ accurately. For example in BaFe$_{1.85}$Co$_{0.15}$As$_2$,
$\w_{\rm res}$ is observed as $10$meV in neutron scattering measurement.\cite{keimer}
In this material, $2\Delta=\Delta_{\rm h}+\Delta_{\rm e}$, where $\Delta_{\rm h(e)}$ denotes magnitude of gap on the hole
(electron) pocket. It was estimated as $11$meV according to an earlier ARPES measurement.\cite{Terashima}
However, $\Delta_{\rm h}+\Delta_{\rm e}$ was estimated as $7$meV by a recent measurement of the specific
heat\cite{Hardy}. We also obtain $\Delta_{\rm h}+\Delta_{\rm e}=7$meV
from a recent penetration depth measurement in Ref. \cite{Luan}, by the
linear interpolation for $x=0.14$ and $x=0.17$.

In our previous paper\cite{onari-resonance}, 
we revealed that for $\Delta=50$meV a prominent hump structure 
{\it free from the resonance mechanism}
appears in neutron scattering spectrum just above $2\Delta$
in the $s_{++}$-wave state.
The hump structure originates from the dissipationless quasi-particles (QPs) 
free from the inelastic scattering in the SC state.
Although the broad spectral peak observed in iron pnictides 
was naturally reproduced based on the $s_{++}$-wave state,
rather than the $s_\pm$-wave state,
used model parameters were not realistic.

In this paper, we investigate the dynamical spin susceptibility 
$\chi^{\rm s}(\w,\Q)$ based on the five-orbital model \cite{Kuroki}
for both $s_{++}$- and $s_\pm$-wave states,
by improving the method of numerical calculation.
Using a realistic parameter $\Delta=5$meV, the obtained results
are more realistic than our previous results for $\Delta=50$meV.
\cite{onari-resonance}
In the normal state, $\chi^{\rm s}(\w,\Q)$ is strongly suppressed 
by the inelastic QP damping $\gamma^*$,
which is large due to the strong correlation.
However, this suppression is released in the SC state since the 
inelastic damping $\gamma^*$ disappears for $|\w|\lesssim3\Delta$.
This ``dissipationless mechanism'' induces a hump-shaped enhancement
in $\chi^{\rm s}(\w,\Q)$ in the $s_{++}$-wave state,
just above $2\Delta$ till $\sim3\Delta$.
In the $s_\pm$-wave state, very high and sharp resonance 
peak appears at $\w_{\rm res}<2\Delta$ even in the case of $\Delta=5$meV.
We demonstrate that the broad spectral peak observed in iron pnictides 
is naturally reproduced based on the $s_{++}$-wave state,
rather than the $s_\pm$-wave state. 

In Sec. III.C, we comment that 
Nagai {\it et al.}\cite{Nagai-resonance}
fail to reproduce the spectral gap in the
two-particle Green function $2\Delta$ and are therefore unreliable.
In appendix, we introduce the similar hump structure of the 
neutron scattering spectrum in CeNiSn
much below the Kondo temperature $T_{\rm K}$.
This compound is called Kondo semiconductor since the hybridization gap 
$\Delta$ opens much below $T_{\rm K}$,
while it is an incoherent metal with large inelastic scattering 
above $T_{\rm K}$.
This is another example of the hump structure 
by the ``dissipationless mechanism'', since 
the inelastic scattering is prohibited by the singlet gap $\Delta$.

We note that numerical results are improved from results in the first
version of
preprint\cite{cond-mat}, in which the value of $\gamma^*(\e)$ for
$3\Delta<\e<4\Delta$ was incorrect in our previous numerical calculation.
\section{Formulation}
\subsection{Method of calculation}
Now, we study the $10\times10$ Nambu BCS Hamiltonian ${\hat {\cal H}}_\k$
composed of the five-orbital tight-binding model and 
the band-diagonal SC gap introduced in ref. \cite{onari-impurity}.
Then, the $10\times10$ Green function is given by
\begin{eqnarray}
{\hat {\cal G}}(i\w_n,\k) &\equiv&
\left(
\begin{array}{cc}
{\hat G}(i\w_n,\k) & {\hat F}(i\w_n,\k) \\
{\hat F}^\dagger(i\w_n,\k) & -{\hat G}(-i\w_n,\k) \\
\end{array}
\right)
 \nonumber \\
&=& (i\w_n{\hat 1}-{\hat \Sigma}_\k(i\w_n)-{\hat {\cal H}}_\k)^{-1} ,
 \label{eqn:G}
\end{eqnarray}
where $\w_n=\pi T(2n+1)$ is the fermion Matsubara frequency,
${\hat G}$ (${\hat F}$) is the $5\times5$ normal (anomalous)
Green function, and ${\hat \Sigma}_\k$ is the self-energy in the
$d$-orbital basis. 
In this paper, we assume that the magnitude of the SC gap
is band-independent; $|\Delta^\nu|=\Delta$.

Here, we have to calculate the spin susceptibility 
as function of real frequency.
Numerically, it is rather easy to use the Matsubara frequency method
and the numerical analytic continuation (pade approximation).\cite{maier-scalapino,eremin}
In the present study, however,
we perform the analytical continuation 
before numerical calculation in order to obtain more reliable results.
The irreducible spin susceptibility in the singlet SC state is given by
\cite{takimoto-moriya}
\begin{eqnarray}
\hat{\chi}_{l_1l_2,l_3l_4}^{0{\rm R}}(\omega,\q)
&=&\frac{1}{N}\sum_\k\int\frac{dx}{2}\nonumber\\
& &\left[\tanh\frac{x}{2T}G^{\rm R}_{l_1l_3}(x_+,\k_+)
\rho_{l_4l_2}^{\rm G}(x,\k)\right.\nonumber\\
&&+\tanh\frac{x_+}{2T}\rho^{\rm G}_{l_1l_3}(x_+,\k_+)G^{\rm A}_{l_4l_2}(x,\k)\nonumber\\
&&+\tanh\frac{x}{2T}F^{\rm R}_{l_1l_4}(x_+,\k_+)
\rho^{{\rm F}\dagger}_{l_3l_2}(x,\k)\nonumber\\
&&+\left.\tanh\frac{x_+}{2T} \rho_{l_1l_4}^{\rm F}(x_+,\k_+)
F^{\dagger {\rm A}}_{l_3l_2}(x,\k)\right],
\end{eqnarray}
where $x_+=x+\w$, $\k_+=\k+\q$, 
$l_i=1,\cdots,5$ represents the $d$-orbital,
and A (R) represents the advanced (retarded) Green function.
$\rho_{ll'}^{\rm G}(x,\k)\equiv 
(G_{ll'}^{\rm A}(x,\k)-G_{ll'}^{\rm R}(x,\k))/2\pi i$
and 
$\rho_{ll'}^{\rm F(\dagger)}(x,\k)\equiv 
(F_{ll'}^{(\dagger)\rm A}(x,\k)-F_{ll'}^{(\dagger)\rm R}(x,\k))/2\pi i$
are one particle spectral functions.
Here, we divide $\hat{\chi}^{0{\rm R(A)}}$ into the ``Hermite part'' $\hat{\chi}^0{}'$ and
``non-Hermite part'' $\hat{\chi}^{0}{}''$,
\begin{eqnarray}
\hat{\chi}^{0{\rm R(A)}}
&\equiv&\hat{\chi}^0{}'+(-)i\hat{\chi}^0{}''\nonumber\\
&=&\left[\frac{\hat{\chi}^{0{\rm R}}+\hat{\chi}^{0{\rm
    A}}}{2}\right]+(-)i\left[\frac{\hat{\chi}^{0{\rm
    R}}-\hat{\chi}^{0{\rm A}}}{2i}\right] .
\label{chi0}
\end{eqnarray}
Then, $\hat{\chi}^0{}'$ and $\hat{\chi}^0{}''$ are expressed as
\begin{eqnarray}
\hat{\chi}^0{}'_{l_1l_2,l_3l_4}(\omega,\q)
&=&\frac{\pi}{2N}\sum_\k\int dx\nonumber\\
& &\left[\tanh\frac{x}{2T}\Theta^{\rm G}_{l_1l_3}(x_+,\k_+)
\rho_{l_4l_2}^{\rm G}(x,\k)\right.\nonumber\\
&&+\tanh\frac{x_+}{2T}\rho^{\rm G}_{l_1l_3}(x_+,\k_+)\Theta^{\rm G}_{l_4l_2}(x,\k)\nonumber\\
&&+\tanh\frac{x}{2T}\Theta^{\rm F}_{l_1l_4}(x_+,\k_+)
\rho^{{\rm F}\dagger}_{l_3l_2}(x,\k)\nonumber\\
&&+\left.\tanh\frac{x_+}{2T} \rho_{l_1l_4}^{\rm F}(x_+,\k_+)
\Theta^{{\rm F}\dagger}_{l_3l_2}(x,\k)\right],
\label{Rechi0}
\end{eqnarray}
\begin{eqnarray}
\hat{\chi}^0{}''_{l_1l_2,l_3l_4}(\omega,\q)&=&\frac{\pi}{2N}\sum_\k\int dx\nonumber\\
&&\left[\tanh\frac{x_+}{2T}-\tanh\frac{x}{2T}\right]\nonumber\\
&&\times\left[\rho^{\rm G}_{l_1l_3}(x_+,\k_+)\rho_{l_4l_2}^{\rm
   G}(x,\k)\right.\nonumber\\
&&+\left.\rho^{\rm F}_{l_1l_4}(x_+,\k_+)\rho^{{\rm
    F}\dagger}_{l_3l_2}(x,\k)\right],
\label{Imchi0}
\end{eqnarray}
where we denote $\Theta_{ll'}^{\rm G}(x,\k)\equiv 
(G_{ll'}^{\rm A}(x,\k)+G_{ll'}^{\rm R}(x,\k))/2\pi$
and 
$\Theta_{ll'}^{\rm F(\dagger)}(x,\k)\equiv 
(F_{ll'}^{(\dagger)\rm A}(x,\k)+F_{ll'}^{(\dagger)\rm R}(x,\k))/2\pi$.

We explain that the non-Hermite part satisfies the relation
$\hat{\chi}^{0}{}''(\omega,\q)=0$ for $|\w|<2\Delta$ at $T=0$. Now, we
assume $\w>0$. $\rho_{ll'}^{\rm G,F}(x,\k)=0$ is
given for
$|x|<\Delta$ since the SC gap opens. Then, in order to satisfy
both $\rho_{ll'}^{\rm G,F}(x,\k)\ne0$ and $\rho_{ll'}^{\rm
G,F}(x_+,\k_+)\ne0$, inequalities $|x_+|>\Delta$ and $|x|>\Delta$ are required.
Moreover, 
$\left[\tanh\frac{x_+}{2T}-\tanh\frac{x}{2T}\right]\ne0$ only when 
$x_+\cdot x<0$. 
In order to obtain the finite value of
$\hat{\chi}^{0}{}''(\omega,\q)$ in eq. (\ref{Imchi0}),
all three inequalities must be satisfied for some $x$.
Considering the third inequality, the first two inequalities 
are restricted to
\begin{eqnarray}
x_+&>&\Delta \label{x+},\\
x&<&-\Delta \label{x}.
\end{eqnarray}
They are satisfied for some $x$ only when $\w>2\Delta$.
Therefore, 
$\hat{\chi}^{0}{}''(\omega,\q)\ne0$ for $|\w|>2\Delta$, while
$\hat{\chi}^{0}{}''(\omega,\q)=0$ for $|\w|<2\Delta$.

In the present numerical study, 
we calculate exactly $\hat{\chi}^{0}{}''$ using eq. (\ref{Imchi0}), and calculate
approximately $\hat{\chi}^{0}{}'$ using the Hermite part of eq. (6) in
Ref. \cite{onari-resonance}. Using this method, we can calculate
accurately the imaginary part of the spin susceptibility as we will
discuss later.

Then, the spin susceptibility $\chi^s(\w,\q)$
is given by the multiorbital random-phase-approximation (RPA)
with the intraorbital Coulomb $U$, interorbital Coulomb $U'$, 
 Hund coupling $J$, and pair-hopping $J'$ \cite{Kuroki}:
\begin{equation}
\chi^s(\w,\q)=\sum_{i,j}\left[\frac{\hat{\chi}^{0\rm R}(\omega,\q)}
{1-{\hat S}^0\hat{\chi}^{0\rm R}(\omega,\q)}\right]_{ii,jj},
\label{RPA}
\end{equation}
where vertex of spin channel ${\hat S}^0_{l_1l_2,l_3l_4}=U$, $U'$, $J$ and $J'$ for
$l_1=l_2=l_3=l_4$, $l_1=l_3\ne l_2=l_4$ , $l_1=l_2\ne l_3=l_4$ and
$l_1=l_4\ne l_2=l_3$, respectively.
Hereafter, we put $J=J'=0.15$eV, $U'=U-2J$,
and fix the electron number as 6.1 (10\% electron-doped case).
In the present model, $\chi^s(0,\q)$ takes the maximum value when 
$\q$ is the nesting vector $\Q=(\pi,\pi/8)$. Due to the nesting,
$\chi^s(0,\Q)/\chi^0(0,\Q) \approx 1/(1-\a_{\rm St})$ is enhanced;
$\a_{\rm St}\ (\lesssim1)$ is 
the maximum eigenvalue of ${\hat S}^0\hat{\chi}^{0\rm R}(0,\Q)$
that is called the Stoner factor.

In the following, we prove that the non-Hermite part of spin susceptibility 
${\rm Im}\chi^s(\w,\q)\equiv [{\chi}^{s\rm R}(\w,\q)-{\chi}^{s\rm A}(\w,\q)]/2i$ 
is zero for $|\w|<2\Delta$ at $T=0$, except at the resonance energy
$\w_{\rm res}$ for the $s_\pm$-wave state:
The spin susceptibility is expressed as
${\chi}^{s\rm R(A)}(\w,\q)=\sum_{l,m}[\hat{\chi}^{s\rm R(A)}]_{ll,mm}$,
where $\hat{\chi}^{s\rm R(A)}\equiv\hat{\chi}^{0\rm R(A)}[1-{\hat
S}^0\hat{\chi}^{0\rm R(A)}]^{-1}$. 
As explained, $\hat{\chi}^0{}''=0$ is satisfied for $\w<2\Delta$.
Then, we obtain $\hat{\chi}^{s\rm R}=\hat{\chi}^{s\rm
A}=\hat{\chi}^0{}'[1-{\hat S}^0\hat{\chi}^0{}']^{-1}$. As a result
$\hat{\chi}^{s\rm R}-\hat{\chi}^{s\rm A}=0$
for $\w<2\Delta$ except when $\det[1-{\hat S}^0\hat{\chi}^0{}']=0$,
which is satisfied at $\w=\w_{\rm res}$ in the $s_\pm$-wave state.
Thus, if we perform the numerical calculation of 
Eqs. (\ref{chi0})-(\ref{RPA}) accurately, Im$\chi^s(\w,{\bm Q})=0$ 
should be satisfied for $\w<2\Delta$.


\subsection{Inelastic QP damping rate $\gamma^*$}
In strongly correlated systems, 
$\chi^s(\w,\q)$ is renormalized by the self-energy correction.
We phenomenologically introduce a band-diagonal self-energy as
$z \cdot {\rm Im}{\hat \Sigma}_\k^{\rm R}(\e)= i\gamma^*(\e){\hat 1}$,
where $z\equiv m/m^*$ is the renormalization factor.
First, we estimate the QP damping in the normal state from
the experimentally observed conductivity.
From the Nakano-Kubo formula, the conductivity is given by 
$\s=e^2\sum_\nu N_\nu(0)v_{\nu}^2/2\gamma(0)$,
where $\gamma(0)\equiv\gamma^*(0)/z$ is the ``unrenormalized'' damping
at zero energy, and $N_\nu(0)$ and $v_\nu$ are the density of states (DOS) and
the Fermi velocity of the $\nu$-th FS, respectively.
Using the five-orbital model,
we obtain $\rho\approx(2.0\gamma(0) {\rm [meV]})$ $\mu\Omega$cm for the
inter-layer spacing $c=6${\AA}  and $\rho\approx(2.8\gamma(0) {\rm [meV]})$
$\mu\Omega$cm for $c=8${\AA}. \cite{onari-impurity,kontani-sato}
In table \ref{table1}, we show the $T$-dependence of $\rho$ 
estimated by fitting the experimental data below $\sim100$K
\cite{Ong,Sefat1,Sefat2},
and the inelastic damping $\gamma(0)$ is derived from 
the theoretical relation between $\rho$ and $\gamma$.
For example, in BaFe$_{1.85}$Co$_{0.15}$As$_2$ ($c\approx6${\AA}),
the unrenormalized inelastic damping $\gamma(0)$ 
is estimated as $3.7 T$,
which is comparable to that in over-doped cuprates.
%
\begin{table*}[t]
\begin{center}
{
\begin{tabular}{|c|c|c|c|}
\hline
 &
 Ba$_{1-x}$K$_x$Fe$_2$As$_2$ [$T_{\rm c}=37$K]\cite{Ong}
&BaFe$_{1.85}$Co$_{0.15}$As$_2$ [$T_{\rm c}=25$K]\cite{Sefat1}&
 LaFeAsO$_{0.89}$F$_{0.11}$ [$T_{\rm c}=28$K]\cite{Sefat2}\\
\hline
$\rho(T)-\rho(0)$[$\mu\Omega$cm]& $\sim23T$& $\sim7.3T$ &
 $\sim4.6T^2$\\
\hline
$\gamma(0)$[meV] at $T$&
 $\sim12T$& $\sim3.7T$ & $\sim 1.6T^2$\\
\hline
$\gamma(0)$[meV] at $T_{\rm c}$& $\sim 37$ & $\sim 7.9$ &
 $\sim 9.3$\\
\hline
\end{tabular}
}
\caption{$\rho(T)-\rho(0)$ and ``unrenormalized''
inelastic damping at zero energy $\gamma(0) \ (=\gamma^*(0)/z)$ 
estimated by fitting the experimental data below $\sim100$K. 
\cite{Ong,Sefat1,Sefat2}.
The unit of $T$ is [meV].
}
\label{table1}
\end{center}
\end{table*}

Then, we derive the ($\e$, $T$)-dependences of the 
``renormalized'' inelastic scattering:
In the presence of the strong spin and orbital fluctuations, the damping
follows the approximate relation $\gamma^*(\e)\approx b(T+|\e|/\pi)$
according to spin (orbital) fluctuation theories. \cite{onari-flex,Pines} 
According to Table \ref{table1},
we obtain $b\sim1.9$ in BaFe$_{1.85}$Co$_{0.15}$As$_2$ if we assume $z\sim0.5$.
In the present study, we use a larger value $b=2.5$.
Note that the result is not so sensitive to the value of $b$.


In the present numerical study, 
we assume more simple $\e$-dependence of $\gamma^*(\e)$ 
to simplify the analysis,
justified in calculating Im$\chi^s$ for $0\le|\w|\lesssim4\Delta$.
In the normal state, we put
\begin{eqnarray}
\gamma^*(\e)=\gamma^*_0.
 \label{gamma0}
\end{eqnarray}
%
In the SC state at $T\ll T_c$, $\gamma^*(\e)=0$ for $|\e|<3\Delta$ 
($=$ a particle-hole excitation gap ($2\Delta$) plus a 
single-particle excitation gap ($\Delta$)),
while its functional form approaches to that of the normal state
for $|\e|\gtrsim3\Delta$. 
Taking these facts into account, we put
\begin{eqnarray}
\gamma^*(\e)=a(\e)\gamma^*_s ,
 \label{a}
\end{eqnarray}
where (i) $a(\e)\ll1$ for $|\e|<3\Delta$,
(ii) $a(\e)=1$ for $|\e|>4\Delta$, and
(iii) linear extrapolation for $3\Delta<|\e|<4\Delta$;
see Fig. \ref{Schema}.
We have confirmed that the obtained results are insensitive to the boundary of 
$|\e|$ ($4\Delta$ in the present case) between (ii) and (iii).
Since $\gamma^*(\e)$ is an increase function of $T$, 
$\gamma^*_s$ at $T\ll T_{\rm c}$ should be smaller than $\gamma^*_0$. 
Here, we derive the values of $\gamma^*_s$ and $\gamma^*_0$
from the relations
$\gamma^*(\e)\sim2.5(T+|\e|/\pi)$,
by putting $\e=3\Delta=15$meV since we are interested in 
the hump structure around $\w\sim3\Delta$.
Therefore, we put $\gamma^*_0=\gamma^*(3\Delta)=20$meV 
at $T=3$meV in the normal state.
Similarly, we put $\gamma^*_s=\gamma^*(3\Delta)=10$meV 
at $T=0$ in the superconducting state.
In the $s_{++(\pm)}$-wave state, we put
$\Delta=5{\rm meV}$ for the two hole-pockets and
$\Delta=(-)5{\rm meV}$ for electron-pockets.
In the numerical calculation, we use $3072\times3072$ $\k$-meshes and 
$a(\e)\gamma_s^*=0.5{\rm meV}\ (=0.1\Delta)$ for $|\e|<3\Delta$.

\subsection{Hump structure in Im$\chi^s$ due to 
dissipationless QPs ($E_\k<3\Delta$)}

\begin{figure}[htb]
\includegraphics[width=0.8\linewidth]{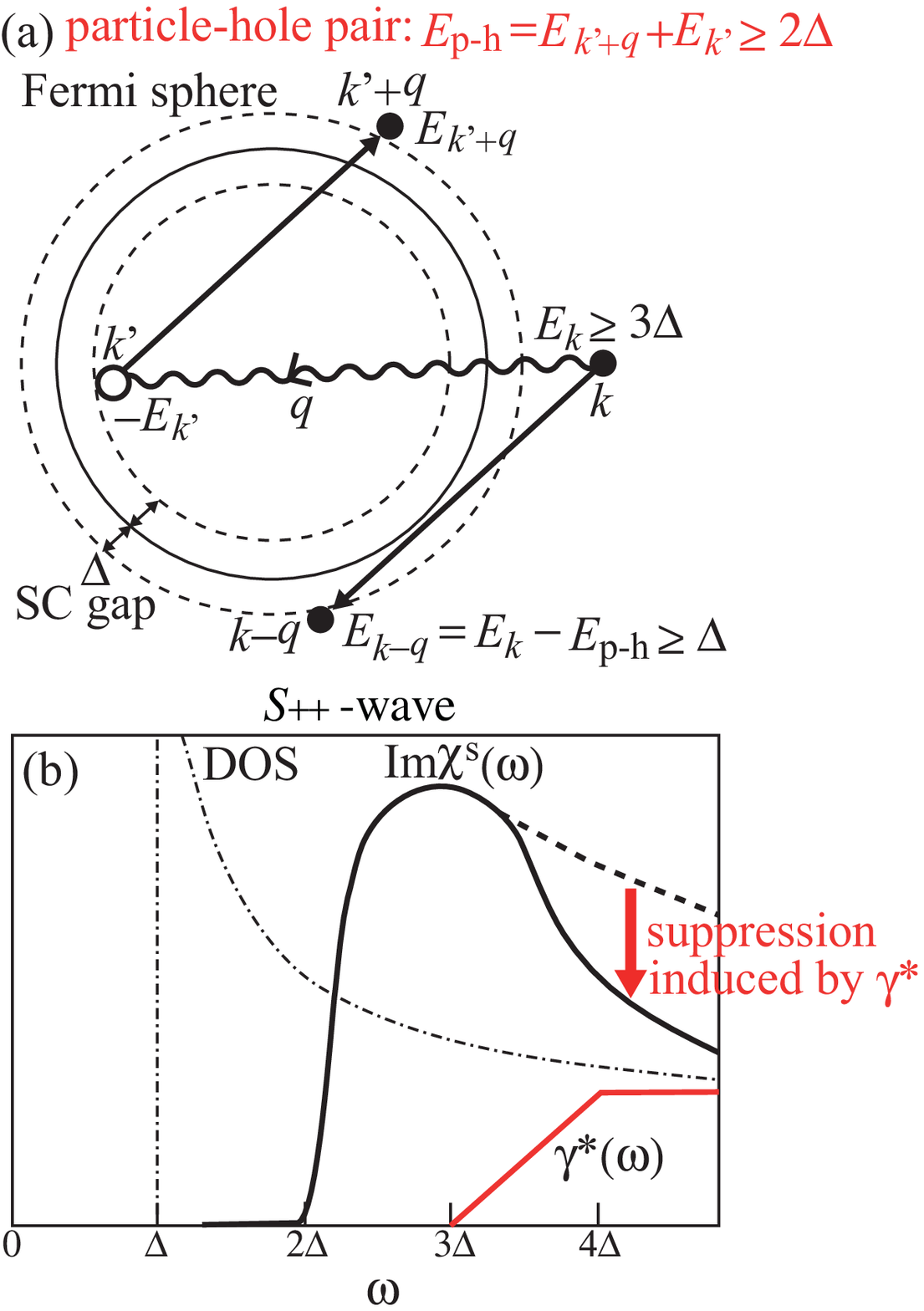}
\caption{
(Color online) 
(a) Schematic inelastic scattering process in the SC state, 
by creating a particle-hole excitation $2\Delta$.
The realization condition is $E_\k\ge 3\Delta$.
(b) Energy dependences of the DOS, $\gamma^*(\omega)$, and 
Im$\chi^s(\omega)$ in the $s_{++}$-wave state.
}
\label{Schema}
\end{figure}
Here, we explain an intuitive reason why 
the QP is ``dissipationless'' for $|\w|<3\Delta$ 
at zero temperatures. \cite{chubukov-resonance}
In Fig. \ref{Schema} (a), 
we show an inelastic scattering process, in which a QP at $\k$ is
scattered to $\k-\q$, with exciting a particle-hole (p-h) pair ($\k'+\q$, $\k'$).
Since a QP in the SC state cannot exist in the thin shell $|\w|<\Delta$,
the particle-hole excitation energy $E_{\rm p-h}$ is always larger than
$2\Delta$. 
Since the energy of the final state $E_{\k-\q}$ is also larger than
$\Delta$, the inelastic scattering is prohibited when $E_{\k}\le 3\Delta$.
Thus, the relationship $\gamma^*(\w)=0$ for $|\w|<3\Delta$ is obtained.
Form this relation, the peak of the DOS at $\w=\Delta$ 
for the isotropic SC gap remains to be sharp.

Then, the dissipationless QPs in the SC state
produce the hump-shaped enhancement in the spin spectrum.
In the normal state, Im$\chi^s$ has no gap structure, and it is
suppressed by the inelastic QP
damping $\gamma^*$ induced by the strong correlation.
In the SC state as illustrated in Fig. \ref{Schema} (b), 
Im$\chi^{\rm s}$ has the p-h excitation gap $2\Delta$.
Since the QP is dissipationless for $|\w|<3\Delta$ in the SC state, 
the suppression in Im$\chi^s(\w)$
is released just above the excitation gap $\w\gtrsim2\Delta$
so as to form a hump structure.
For this reason, a prominent hump structure 
appears in Im$\chi^{\rm s}(\w,\Q)$ just above $2\Delta$ till $\sim3\Delta$
in strongly correlated $s_{++}$-wave superconductors.

\section{Numerical Result}
\subsection{Spin susceptibility at the nesting vector $Q=(\pi,\pi/8)$}
\begin{figure}[htb]
\includegraphics[width=0.76\linewidth]{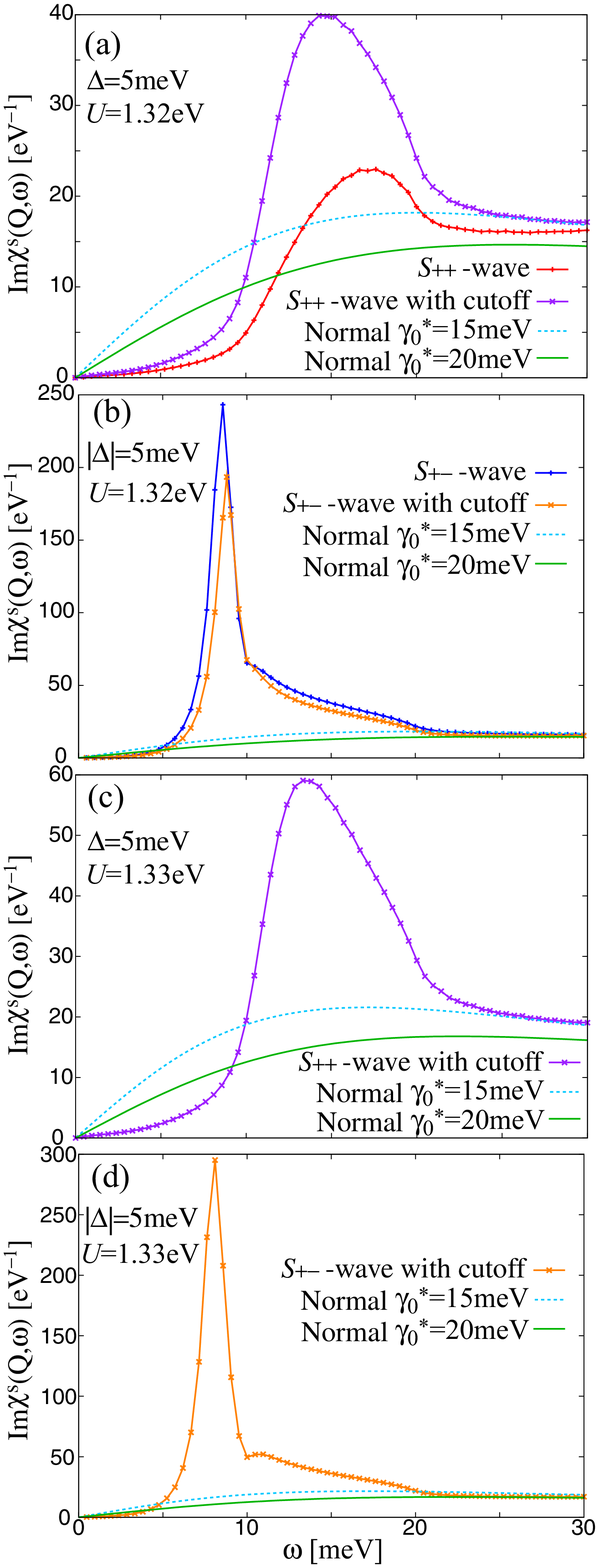}
\caption{
(Color online) 
(a) $\w$-dependence of Im$\chi^s(\w,{\bm Q})$ at ${\bm Q}=(\pi,\pi/8)$
 for $U=1.32$eV in the $s_{++}$-wave state ($\Delta=5$meV),
as well as in the normal state with $\gamma^*_0=15$, $20$meV.
The hump structure is enhanced by considering the high-energy
dependence of the SC gap, by introducing the cutoff energy $\Delta E=20$meV,
(b) those in the $s_{\pm}$-wave state
 ($|\Delta|=5$meV), (c) those for $U=1.33$eV in the $s_{++}$-wave state,
 and (d) those in the
 $s_{\pm}$-wave state.
}
\label{Fig1}
\end{figure}

Figure \ref{Fig1} shows the obtained 
Im$\chi^s(\w,\Q)$ at the nesting vector between
the hole- and electron-pockets $\Q=(\pi,\pi/8)$:
We fix $T=1$meV hereafter, since the obtained results are
insensitive to the temperature for $T\le 3$meV.
In the normal state with $\gamma^*_0=20(15)$meV, the Stoner factor is 
$\a_{\rm St}=0.950(0.959)$ for $U=1.32$eV.
In the SC states with $\gamma^*_s=10$meV, $\a_{\rm St}=0.956$ $(0.982)$ in the 
$s_{++}$-wave ($s_{\pm}$-wave) state for $U=1.32$eV.
In the $s_{\pm}$-wave state, $\a_{\rm St}$ increases due to the
coherence factor.
Inversely, $\a_{\rm St}$ in the $s_{++}$-wave state
decreases due to absence of coherence factor.
As shown in Fig. \ref{Fig1}, in the normal state with
$\gamma^*_0=20$meV, the peak position of Im$\chi^s$ 
is about $20-25$meV, which is consistent
with experimental result in BaFe$_{1.85}$Co$_{0.15}$As$_2$
\cite{keimer}.
Thus, the value of Im$\chi^s$ in the normal state with
$\gamma^*_0=15$meV is overestimated. 

A broad hump structure appears in the $s_{++}$-wave state at
$\omega\gtrsim 2\Delta$ even in the case of
$\Delta=5$meV, and its overall shape is consistent with experimental
results\cite{qiu,keimer}.
We had neglected the energy-dependence of $\Delta$ in the previous study
 \cite{onari-resonance}.
However, in reality, the SC gap $\Delta$ 
will be cut off when the energy of the $\nu$-th
band $\e^\nu_{\k}$ measured from the
Fermi energy exceeds the characteristic energy scale of the pairing interaction.
To take this fact into account, we introduce a Gaussian cutoff 
$\Delta^\nu_{\k}=\Delta^\nu\exp\{-[\e^\nu_{\k}/\Delta E]^2\}$ following
Refs. \cite{Maier,Nagai-resonance}.
We put $\Delta E=20$meV, which correspond to the Fe ion optical phonon
frequency $\omega_{\rm D}\sim 20$meV employed in the orbital
fluctuation theory.\cite{Kontani-Onari,onari-flex}.

When the cutoff is applied in the $s_{++}$-wave state, the hump structure becomes more
prominent as shown in Fig. \ref{Fig1} (a). We confirm that the obvious hump
appears over the 
normal state even with $\gamma^*_0=15$meV.
The enhancement of hump structure originates from the increment of the 
Stoner factor by introducing the cutoff, 
from $\a_{\rm St}=0.956$ to $0.965$ for $U=1.32$eV. 

On the other hand, in the $s_{\pm}$-wave state, very high and sharp resonance peak
appears at $\omega_{\rm res}<2\Delta$ even in the case of
$|\Delta|=5$meV as shown in Fig. \ref{Fig1} (b). This result is
apparently inconsistent with experimental results. 
In order to explain the experimental result by the $s_{\pm}$-wave state,
large inhomogeneity would be required, although the $s_{\pm}$-wave state
is fragile against inhomogeneity.
The height of the resonance peak exceeds $100$eV$^{-1}$ for $a(0)\gamma^*_s=0.5$meV,
while it diverges for $a(0)\rightarrow0$ if $\k$-meshes are fine enough.
Im$\chi^s$ is slightly suppressed by considering the cut off, $\Delta E=20$meV.

We also study the spectra for both $s_{++}$- and $s_\pm$-wave states 
with cutoff for $U=1.33$eV:
In Fig. \ref{Fig1}(c) and \ref{Fig1}(d), we show the results for 
the normal state with $\gamma^*_0=15$meV ($\a_{\rm St}=0.965$) and $\gamma^*_0=20$meV ($\a_{\rm St}=0.956$).
We also show results for the $s_{++}$-wave state with $\gamma^*_s=10$meV ($\a_{\rm St}=0.971$), and
$s_\pm$-wave state with $\gamma^*_s=10$meV ($\a_{\rm St}=0.984$).

We note that the effect of multiband on Im$\chi^s$, 
which was discussed in Ref. \cite{keimer}, 
is automatically included in our calculation.
By increasing $U$ from $1.32$eV to $1.33$eV,
the hump structure in the $s_{++}$-wave state is more enhanced.
Also, the resonance peak in the $s_\pm$-wave state develops,
and $\w_{\rm res}$ shifts to lower energy.

\begin{figure}[htb]
\includegraphics[width=0.85\linewidth]{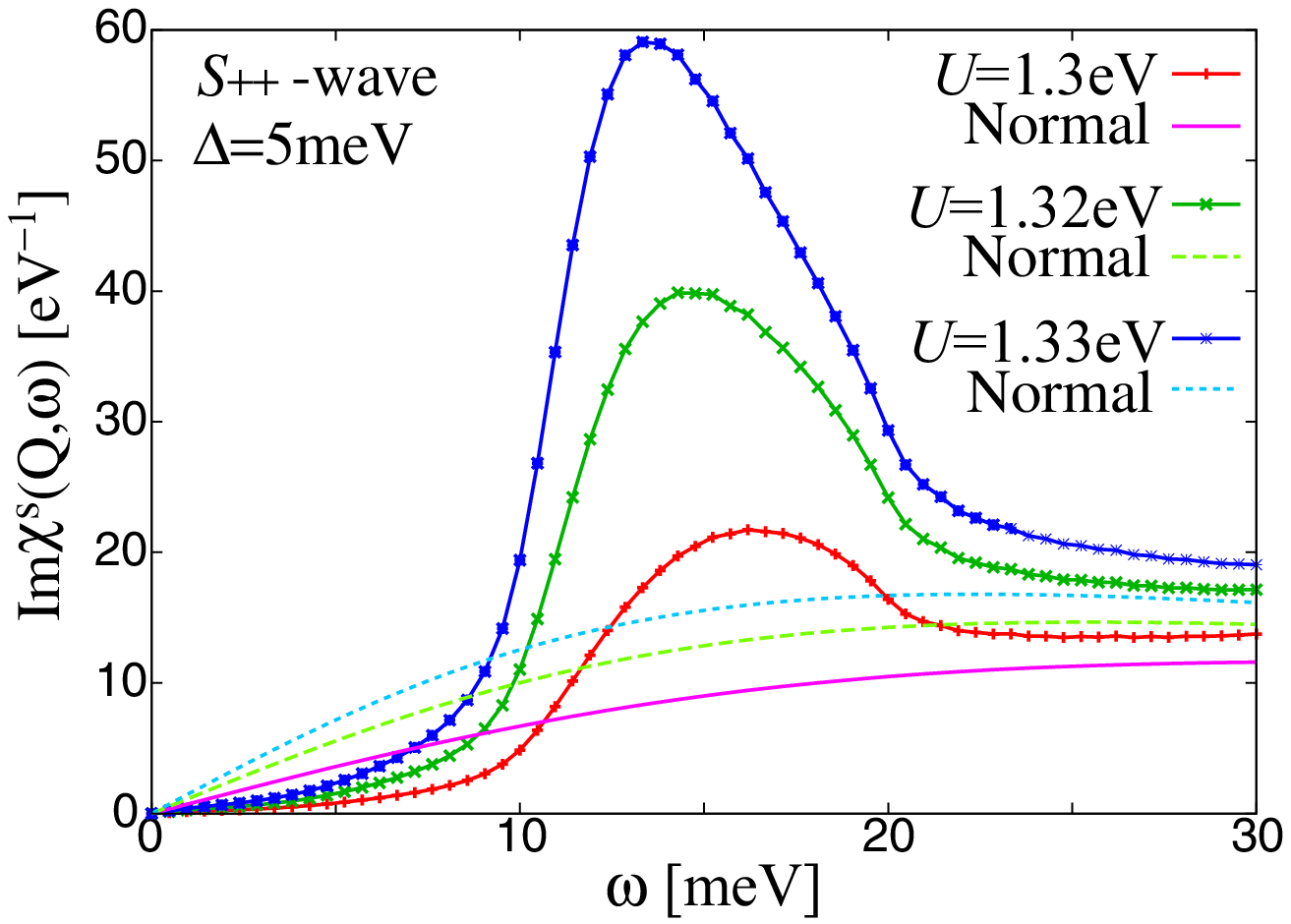}
\caption{
(Color online) 
$\w$-dependence of Im$\chi^s(\w,{\bm Q})$ at ${\bm Q}=(\pi,\pi/8)$ for
$U=1.3$, $1.32$, and $1.33$eV in the $s_{++}$-wave state with $\Delta=5$meV, and the cutoff
 energy $\Delta E=20$meV, as well as normal states with
 $\gamma_0^*=20$meV for each value of $U$.
}
\label{U132}
\end{figure}
In Figure \ref{U132}, we confirm that the hump in the $s_{++}$-wave
with $\Delta=5$meV state is enhanced as the value of $U$ increases.
Thus, the hump becomes prominent as system comes close to the AF order.
In this paper, we have calculated $\chi^s$ introduced in
Eq. \ref{RPA}. To obtain the value of spin susceptibility $\chi^{\rm
neu}$ observed in neutron measurements, we have to take the spin
magnetic moment $(=1\mu_{\rm B})$ and the factor of spin degeneracy.
Its $z$-component is $\chi_z^{\rm
neu}=2\chi^s$[$\mu_{\rm B}^2$eV$^{-1}$] and the transverse spin
susceptibility is $\chi_{\pm}^{\rm
neu}=4\chi^s$[$\mu_{\rm B}^2$eV$^{-1}$].

\subsection{Comparison with our previous method}
\begin{figure}[htb]
\includegraphics[width=0.8\linewidth]{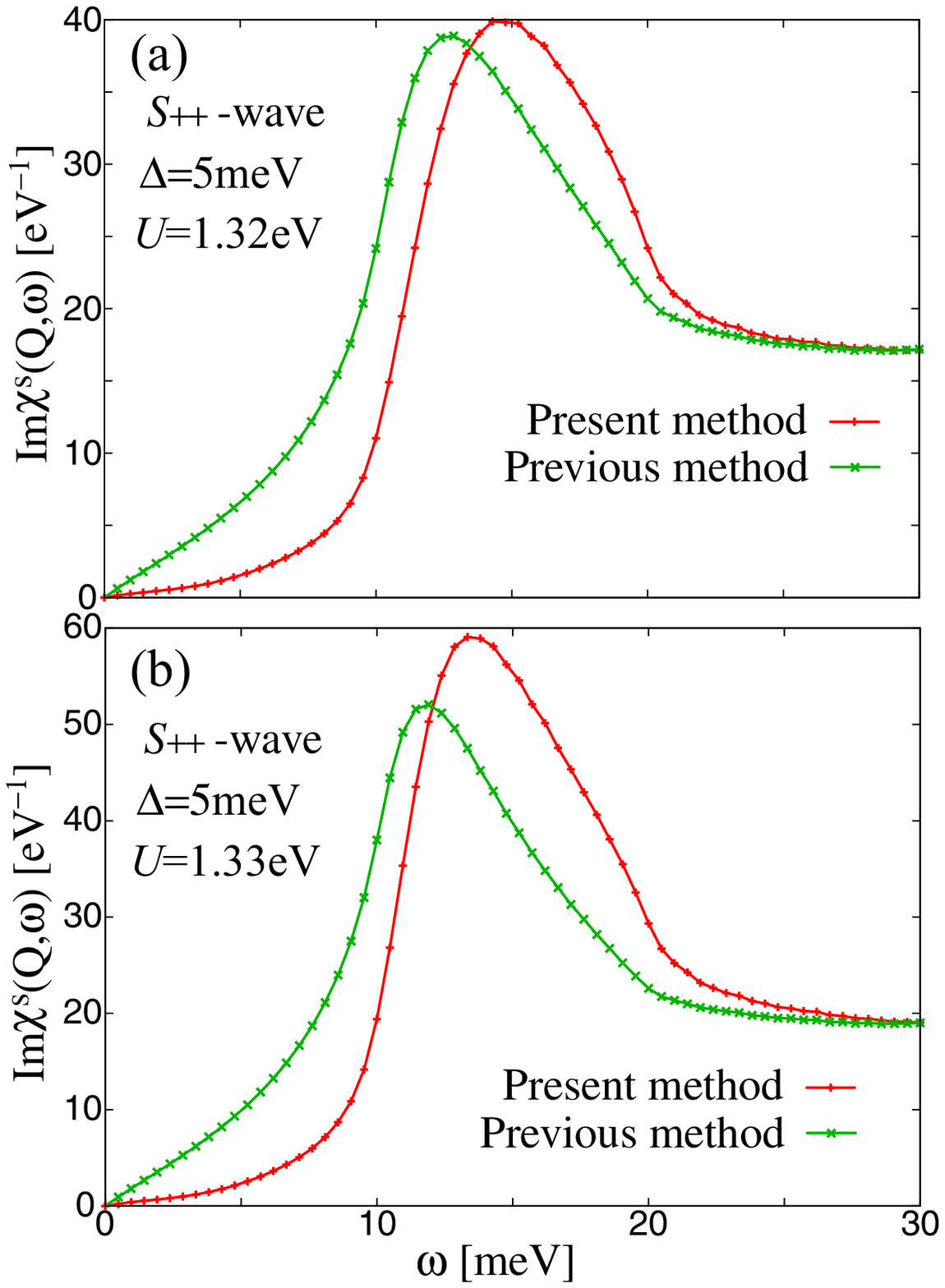}
\caption{
(Color online) 
Im$\chi^s(\w,{\bm Q})$ at ${\bm Q}=(\pi,\pi/8)$
for $U=1.32$eV (a) and $U=1.33$eV (b) in the $s_{++}$-wave state ($\Delta=5$meV).
We show the comparison between the present improved method 
and the previous method in Ref. \cite{onari-resonance},
with the cutoff ($\Delta E=20$meV).
}
\label{Fig3}
\end{figure}

In the $s_{++}$-wave state,
Im$\chi^s(\w,{\bm Q})=0$ for $|\w|<2\Delta$ at $T=0$
as we discussed in Sec. II.
This relation is correctly satisfied in the present method
if we put $a(0)\rightarrow0$ in Eq. (\ref{a}).
In the present method,
we perform the numerical calculation of ${\hat \chi}^0{}''$ 
using Eq. (\ref{Imchi0}) exactly.
In fact, in Fig. \ref{Fig3}(a) and (b), we verify that the spectral gap of
Im$\chi^s(\w,{\bm Q})$ is well reproduced 
in the present method with $a(0)\gamma_s^*=0.5$meV, 
demonstrating the superiority of the present method
to the previous method in Ref. \cite{onari-resonance}.
In the case of $s_\pm$-wave state,
we obtain Im$\chi^s(\w,{\bm Q})\propto\delta(\w-\w_{\rm res})$ for
$|\w|<2\Delta$ if the numerical calculation is performed accurately.

In the present paper, we calculate $\hat{\chi}^0{}''$ in
eq. (\ref{Imchi0}) exactly, while
$\hat{\chi}^0{}'$ is calculated 
approximately using eqs. (6) and (7) in Ref.\cite{onari-resonance}.
We consider this is justified since we had verified that the present 
``approximated RPA'' is reliable in our
previous paper\cite{onari-resonance}:
In Fig. 1 (b) of Ref.\cite{onari-resonance}, we
had performed the ``exact RPA calculation'' for both $\hat{\chi}^0{}'$ and
$\hat{\chi}^0{}''$ with $\Delta=400$meV, and confirmed that overall behavior
of Im$\chi^s(\w,{\bm
Q})$ is well reproduced by the present approximated RPA.

Here, we comment on the $\bm{q}$ dependence of Im$\chi^s(\w,{\bm q})$
around $\bm{q}=\bm{Q}$.
In our two-dimensional model\cite{Kuroki}, it is difficult to discuss the $\bm{q}$
dependence of Im$\chi^s(\w,{\bm q})$ because $\bm{q}$ dependence of
Im$\chi^s(\w,{\bm q})$ is drastic even in the normal state, which is
inconsistent with the neutron scattering measurements.


\subsection{Comparison with Nagai {\it et al.}\cite{Nagai-resonance}}
\begin{figure}[htb]
\includegraphics[width=\linewidth]{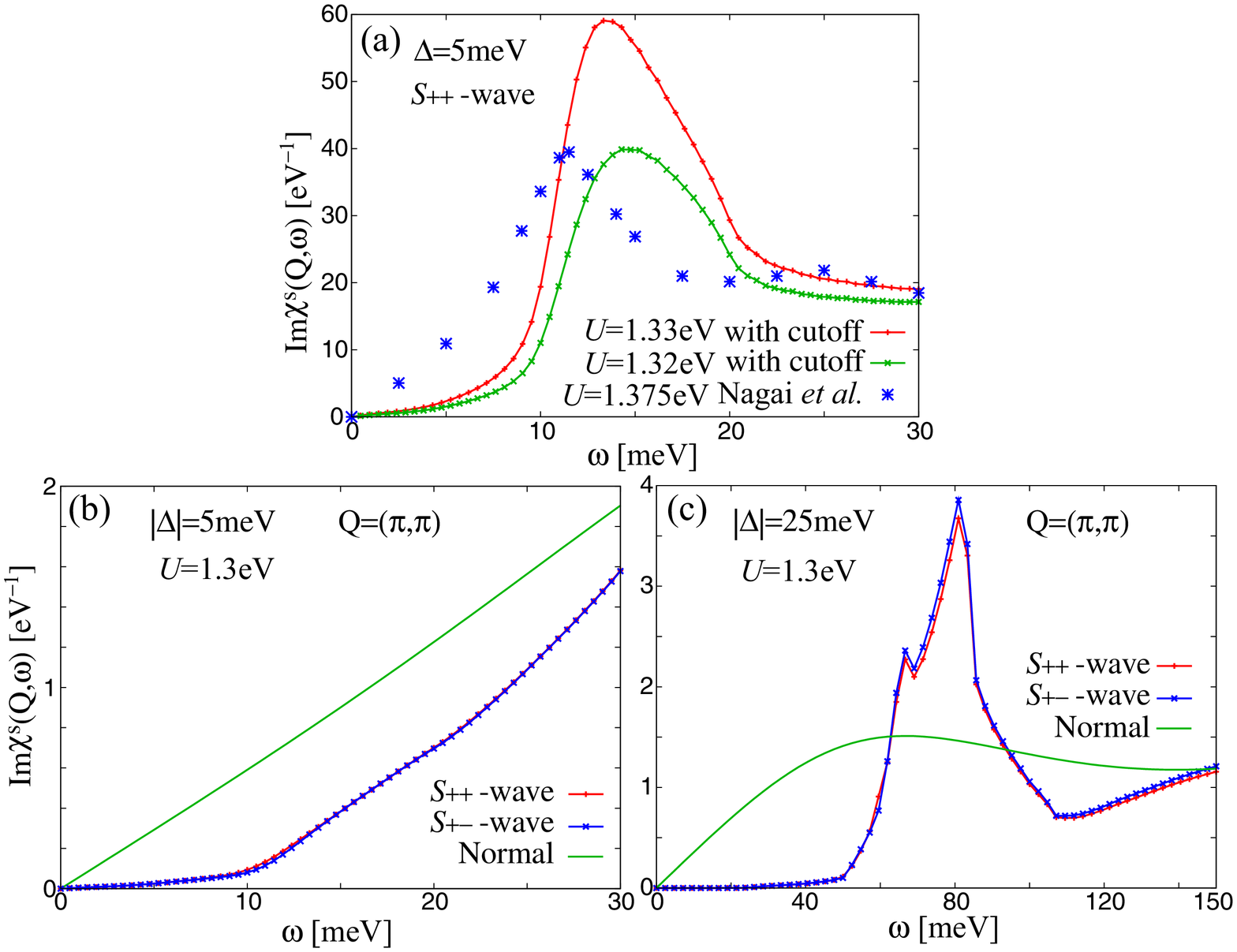}
\caption{
(Color online) (a) Im$\chi^s(\w,{\bm q})$ in
the $s_{++}$-wave state ($\Delta=5$meV) for $U=1.32$eV and $1.33$eV
obtained in the present study.
We also plot the data of Nagai {\it et al.}\cite{Nagai-resonance} for $U=1.375$eV,
by multiplying 0.39.
All results are obtained for the cutoff energy $\Delta E=20$meV.
(b) Im$\chi^s(\w,{\bm q})$ at ${\bm q}=(\pi,\pi)$
for $U=1.3$eV in both $s_{++}$- and $s_{\pm}$-wave states
with $|\Delta|=5$meV and $\gamma^*_s=10$meV,
together with the result in the normal state for $\gamma^*_0=20$meV.
(c) Im$\chi^s(\w,{\bm q})$ at ${\bm q}=(\pi,\pi)$
in both $s_{++}$- and $s_{\pm}$-wave states with $|\Delta|=25$meV
and $\gamma^*_s=50$meV, together with the result in the normal state for
$\gamma^*_0=50$meV.
}
\label{Fig2}
\end{figure}
Recently, Nagai {\it et al.}\cite{Nagai-resonance} had calculated the neutron
scattering spectrum using the method proposed in Ref.\cite{onari-resonance}, and 
claimed that (i) hump structure in the
$s_{++}$-wave state is smeared when $\gamma^*\sim 10$meV and $\Delta\sim 5$meV 
compared to the case of $\Delta>25$meV, and (ii) resonance peak in the
$s_{\pm}$-wave state becomes very low and broad.
Moreover, they had also claimed that
(iii) one can distinguish
between the $s_{++}$-wave and the $s_{+-}$-wave states from the
spectrum at $\q=(\pi,\pi)$.

First, we explain that (i) and (ii) are incorrect statements based on
their inaccurate numerical calculation. First, their result fails to
reproduce the spectral gap of Im$\chi^s$ for $\w<2\Delta$ as shown in
Fig. \ref{Fig2} (a). 
(One can prove rigorously that Im$\chi^s=0$ for $\w<2\Delta$ at $T=0$.)
Second, the peak position of the result of Nagai {\it et
al.}\cite{Nagai-resonance} is about $2\Delta$, while it must be higher
energy ($\sim 3\Delta$). 
In the $s_\pm$-wave state, the resonance peak
should be $\delta$ functional structure when
$a(0)$ in eq. (\ref{a}) is enough smaller than $\Delta$.
Thus, the low and broad resonance peak of Nagai {\it et
al.}\cite{Nagai-resonance} is far from the exact
behavior of the resonance peak.
In Nagai's results, fine structures in Im$\chi^s(\w,\q)$ 
seem to be inappropriately smeared
in both $s_{++}$- and $s_\pm$-wave states.



Next, we comment on the claim (iii). 
They pointed out the spectrum in the $s_\pm$-wave state with $\Delta=5$meV
are different from that in the $s_{++}$-wave state with $\Delta=25$meV.
Here, we show the results of both $s_{++}$- and $s_\pm$-wave states in
Figs. \ref{Fig2}, for (b) $|\Delta|=5$meV and (c) $|\Delta|=25$meV.
Since both spectra are almost identical, we cannot
distinguish between the $s_{++}$- and $s_{\pm}$-wave states by the
spectrum at the wave vector $\q=(\pi,\pi)$ {\it for the same $\Delta$}.
This result is reasonable
because sign of the SC gap is preserved through the $(\pi,\pi)$ shift 
for both the $s_{++}$- and $s_{\pm}$-wave states.
Although claim (iii) is based on their numerical result in
which the hump of the $s_{++}$-wave state appears only for
$\Delta\gtrsim25$meV, the prominent hump appears in the $s_{++}$-wave state with
$\Delta=5$meV in our improved numerical results as shown in
Fig. \ref{Fig1}(a) and (c). Thus, we conclude it is impossible to
distinguish between the $s_{++}$- and $s_\pm$-wave states with $\Delta=5$meV.
\section{Conclusion}
We have studied the dynamical spin susceptibility $\chi^{\rm s}(\w,\Q)$ 
in iron-based superconductors
for both $s_{++}$- and $s_\pm$-wave states, by developing more accurate 
numerical method and introducing the high-energy dependence of the SC gap.
\cite{onari-resonance}
In the $s_{++}$-wave state, the dissipationless QPs for $|\w|<3\Delta$
produce a prominent hump-shaped enhancement in $\chi^{\rm s}(\w,\Q)$ 
just above $2\Delta$ till $\sim3\Delta$.
This ``dissipationless mechanism'' is unrelated to the resonance.
The peak energy of the hump will shift to lower energy
if we consider the band-dependence and/or the anisotropy of the SC gap,
as we discussed in Ref. \cite{onari-resonance}.

On the other hand, in the $s_\pm$-wave state, very high and sharp resonance 
peak appears at $\w_{\rm res}<2\Delta$.
In order to explain small and broad peaks observed in Refs.\cite{qiu,keimer} as the
resonance peak in the $s_\pm$-wave state, sufficient inhomogeneity or small SC
volume fraction would be required. However, the $s_{\pm}$-wave state
is fragile against inhomogeneity.
We concluded that the small and broad spectral peak observed in iron pnictides 
is naturally reproduced based on the $s_{++}$-wave state 
in the absence of inhomogeneity, rather than the $s_\pm$-wave state. 

In the Comment on the present paper written
by Nagai and Kuroki on arXiv\cite{Nagai-resonance2}, the authors repeated
their claim ``smallness of the hump in the $s_{++}$-wave state'' based
on the ``old method'' that was first developed in Ref. \cite{onari-resonance}.
In Sec. III, however, we actually obtained large hump using the ``new
method'', which is mathematically superior to the old method.
This discrepancy originates from the calculation method as well as the
numerical accuracy, not from the detail of model parameters, as we
discussed in our Reply on arXiv\cite{onari-resonance2}.

\acknowledgments
We are grateful to M. Sato, Y. Kobayashi, Y. Matsuda, 
D. S. Hirashima, D. J. Scalapino, P. J. Hirschfeld, A. V. Chubukov,
I. Eremin, Y. Tanaka and K. Kuroki, for valuable discussions. 
This study has been supported by Grants-in-Aid for Scientific 
Research from MEXT of Japan, and by JST, TRIP.
Numerical calculations were performed at the Computer Center 
and the ISSP Supercomputer Center of University
 of Tokyo, and the Yukawa Institute Computer Facility.

\appendix
\section{Hump structure in the neutron inelastic scattering 
for a Kondo semiconductor CeNiSn}
In this paper, we have studied the neutron inelastic scattering
spectrum in iron pnictide superconductors.
In the $s_{++}$-wave SC state, we confirmed that a large hump structure 
appears just above $2\Delta$ due to the reduction in the inelastic
QP scattering $\gamma^*$, which is the most important finding in this paper.

\begin{figure}[htb]
\includegraphics[width=\linewidth]{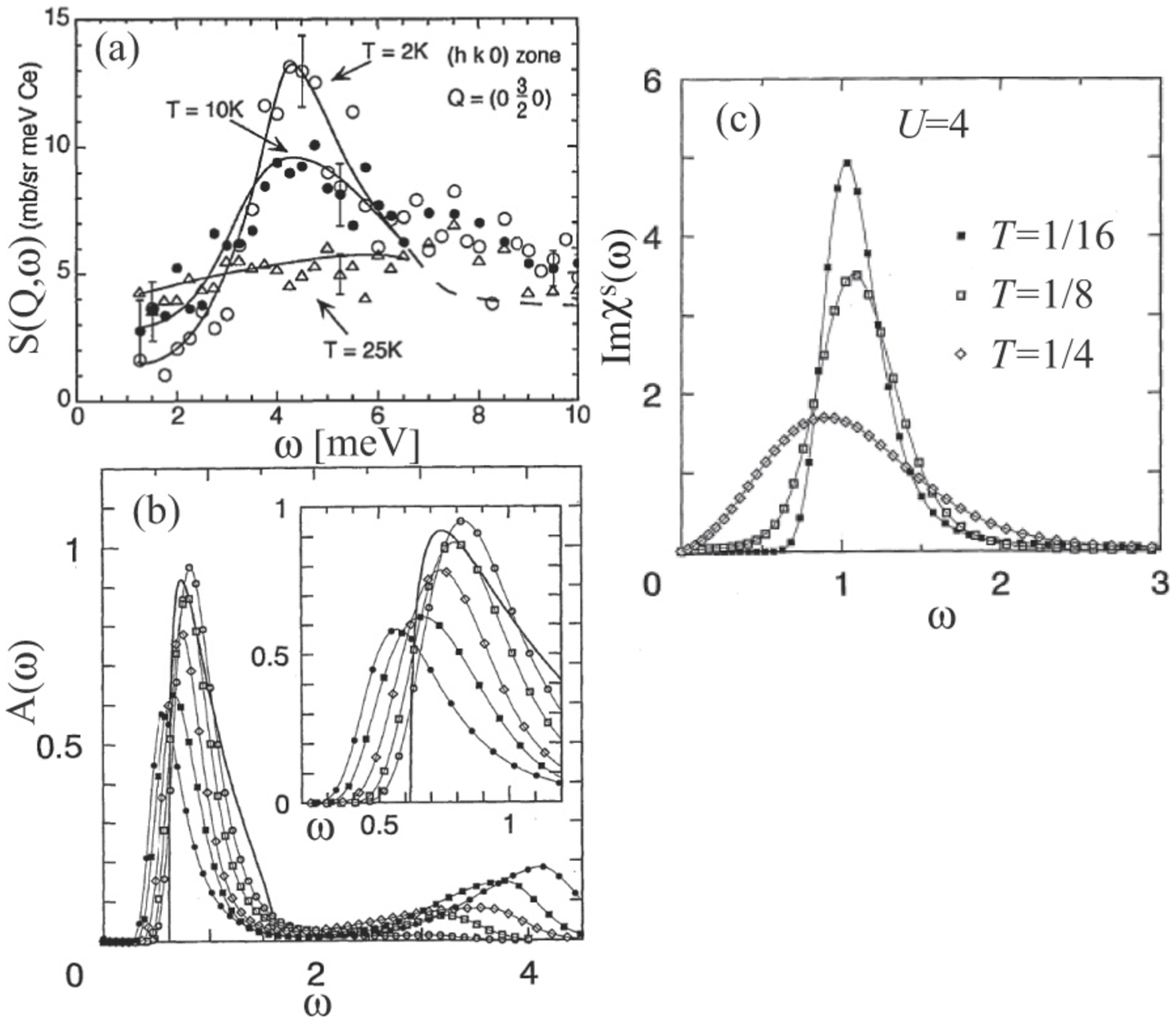}
\caption{ 
(a) $\w$-dependence of $S(\Q,\w)={\rm Im}\chi^s/(1-e^{-\w/T})$ in
 CeNiSn at various temperatures \cite{Kadowaki}.
(b) Single-particle spectrum $A(\w)$ at $T=1/16$ for 
$U=0$ (solid curve), 1(open circles), 2(open squares), 3(open diamonds),
4(solid squares) and 5(solid circles) \cite{mutou-hirashima}.
The inset shows the low frequency part.
(c) Im$\chi^s(\w)$ for $U=4$ at different temperatures
 \cite{mutou-hirashima}.
}
\label{Kondo}
\end{figure}

Then, a natural question is whether such a hump-shaped enhancement
by ``dissipationless mechanism'' is universal or not.
To answer this question, we discuss a Kondo semiconductor CeNiSn.
Figure \ref{Kondo} (a) shows the neutron inelastic scattering 
spectrum in CeNiSn at $\q=(0,\pi,0)$ at low temperatures 
\cite{Kadowaki}.
The observed large and broad hump structure in CeNiSn
\cite{Kadowaki,Raymond} is very similar to that in iron pnictides.
CeNiSn is an incoherent metal with large inelastic scattering
above the Kondo temperature $T_{\rm K}\sim30$K,
while it becomes a semiconductor with $c$-$f$ hybridization gap 
in the single-particle spectrum ($\Delta$) much below $T_{\rm K}$.

The effective model for the CeNiSn is described as the 
periodic Anderson model (PAM) at half-filling. \cite{Ikeda,mutou-hirashima,mutou-hirashima2}
Neglecting the $f$-orbital degeneracy, the PAM is given as
\begin{eqnarray}
{\cal
 H}&=&\sum_{\k,\sigma}\e^c_{\k}c^\dagger_{\k\sigma}c_{\k\sigma}+\e_f\sum_{\k,\sigma}f^\dagger_{\k\sigma}f_{\k\sigma}+U\sum_if^\dagger_{i\uparrow}f_{i\uparrow}f^\dagger_{i\downarrow}f_{i\downarrow}\nonumber\\
&&+V\sum_{\k,\sigma}\left(f_{\k\sigma}^\dagger
					c_{\k\sigma}+c_{\k\sigma}^\dagger
					f_{\k\sigma}\right),
\end{eqnarray}
where $c_{\k\sigma}(c^\dagger_{\k\sigma})$ and
$f_{\k\sigma}(f^\dagger_{\k\sigma})$ are annihilation (creation)
operators for $c$- and $f$-electrons, respectively.
$V$ is the $c$-$f$ mixing potential, and $U$ is the 
Coulomb interaction for $f$-electrons. Here, the bandwidth is $2$.
Mutou and Hirashima studied this model at half-filling
using the dynamical mean-field theory (DMFT) 
and the quantum Monte Carlo (QMC) \cite{mutou-hirashima}.
Hereafter, we introduce their numerical results and discuss the 
energy-dependence of Im$\chi^{\rm s}(\w)$.
Readers can find more detailed explanations in the original paper
\cite{mutou-hirashima}.

Figure \ref{Kondo} (b) shows the obtained single-particle spectrum $A(\w)$.
For $U=0$, the hybridization gap in $A(\w)$ is $\Delta=0.62$.
For $U=4$, $\Delta$ is renormalized to $0.35$ at 
$T=1/16 \ (<T_{\rm K})$, while the gap is smeared out by
thermal fluctuations above $T_{\rm K}$ \cite{mutou-hirashima}.
At $T=0$, inelastic QP scattering is suppressed by the hybridization gap,
such that $\gamma^*(\w)=0$ for $|\w|<3\Delta$\cite{mutou-hirashima2} in
analogy to Fig. \ref{Schema} (a).

Figure \ref{Kondo} (c) shows Im$\chi^{\rm s}(\w)$ for $U=4$.
In the metallic state at $T=1/4\ (\gg T_{\rm K})$,
Im$\chi^{\rm s}(\w)$ shows a gapless metallic behavior.
In the semiconducting state at $T=1/16\ (\ll T_{\rm K})$,
in contrast, it shows a spectral gap $\Delta_s$ 
and the relation $\Delta_s\approx2\Delta\approx0.7$ is recognized.
At the same time, large hump structure emerges 
around $\w\sim 3\Delta$.
Because of the absence of spin resonance mechanism,
its natural explanation is the reduction in the 
inelastic QP scattering ($\gamma^*(\w)=0$ for $|\w|<3\Delta$),
as we discussed in Fig. \ref{Schema} (b).
We must stress the hump structure in Fig. \ref{Kondo} (c)
is obtained {\it exactly in the DMFT}, by including 
both the self-energy and vertex corrections.
Therefore, experimental and theoretical studies in CeNiSn
strongly support the idea of ``hump structure 
in the $s_{++}$-wave state'' given in Fig. \ref{Fig1} (a),
that is obtained by the RPA by introducing the inelastic QP scattering 
$\gamma^*(\w)$ phenomenologically.

\end{document}